\documentstyle[twoside,fleqn,espcrc2]{article}

\newcommand\negsp{\hspace{-0.7em}}
\newcommand{\Journal}[4]{{#1} {\bf #2}, #3 (#4)}

\newcommand\CJP{\em Can. J. Phys.}

\newcommand\PLB{{\em Phys. Lett.}  B}

\newcommand\JMP{\em J.~Math. Phys.}
\newcommand\AP{\em Ann. Phys.}

\newcommand{\AmS}{{\protect\the\textfont2
  A\kern-.1667em\lower.5ex\hbox{M}\kern-.125emS}}

\input epsf.sty

\title{How do constituent quarks arise in QCD?\\ 
       Perturbation theory and the infra-red%
       \thanks{Talk presented by E. Bagan}}

\author{E. Bagan\address{Grup de F{\'{\i}}sica Te{\`o}rica, Departament de 
        F{\'{\i}}sica and IFAE,\\ 
        Edifici Cn, Universitat Aut{\`o}noma de Barcelona \\
        E-08193 Bellaterra (Barcelona) Spain},%
        \addtocounter{address}{-1}
        M. Lavelle\addressmark,
        D. McMullan\address{Department of Mathematics and Statistics, \\
        University of Plymouth, Drake Circus, \\
        Plymouth, Devon PL4 8AA, United Kingdom},%
       \addtocounter{address}{-2}
        B. Fiol\addressmark\ and \addtocounter{address}{-1}
        N. Roy\addressmark}
       
\begin{document}

\begin{abstract}
We motivate the use of dressed charges by arguing that such objects are
needed
to describe, e.g., constituent quarks and, in general, physical charged states in gauge theories. 
%
%
We give a short introduction to dressings in both QED and QCD. We put special emphasis  on the
infra-red properties of a moving dressed charge. To be more precise, we demonstrate that the
one loop
propagator of a relativistic dressed charge can be renormalized in the mass shell scheme 
with no infra-red divergence showing up.
\end{abstract}

\maketitle

\section{WHY DRESSINGS?}

Many aspects of hadronic physics can be well described in terms 
of constituent quarks. The role played by such objects in our 
discovery of colour and QCD is well known. However, we do not yet 
have a good understanding of how such objects can emerge from QCD. 
This talk describes a new approach to this problem (for a review see~\cite{LaMcMu96a}). 

Any description 
of a (colour) charged particle in a field theory has to fulfill certain 
requirements:
({\em i}) we need to include a 
chromo-(electro-) magnetic cloud around the charge. This is known to 
underlie the infra-red problem and we recall the long postulated link 
between the infra-red structure of QCD and confinement; 
({\em ii}) since constituent quarks have, in the realm where the quark model is valid, 
a physical meaning, it is essential to describe them in a gauge invariant 
way. This, in turn, ensures that the constituent quarks have a well defined colour charge as
required by the standard quark model\cite{LaMcMu96a}.

\subsection{QED}

In QED, an approach which incorporates the above requirements
was originally proposed by Dirac\cite{Di55}. If some function $f_\mu(z,x)$ satisfies
$\partial_\mu^z f^\mu(z,x)= 
\delta^{(4)}(z-x)$, then
a charged particle at $x$ with an electromagnetic cloud around it
may be written in a gauge invariant manner as
\begin{equation}
\psi_{\rm f}(x)=\exp\left\{ ie\int d^4zf_\mu(z,x)A^\mu(z)\right\} 
\psi(x)\,.
\label{eq:fcond}
\end{equation}
The phase factor in~(\ref{eq:fcond}) is usually called
the {\em dressing}. There are as many dressings as there are possible choices of $f_\mu$ 
in~(\ref{eq:fcond}).
We will study two such choices. Our first example (which we will refer to as ({\em i}) below)
is
\begin{eqnarray}
&&f_0=f_1=f_2=0,\nonumber\\
&&f_3(z,x)=-\theta(x^3-z^3)\prod_{j=0}^2\delta(z^j-x^j),
\end{eqnarray}
which can be easily seen to correspond to a string attached to the charge at $x$
and going along a straight line in the $x^3$ direction out to infinity. One could equally
well choose a more complicated path $\Gamma$ and in general
we call such a dressed field, $\psi_\Gamma$.
Example ({\em ii}), which is the case we are primarily concerned with, is
\begin{eqnarray}
&&f_0=0,\nonumber\\
&&\vec f ={1\over4\pi}\; \delta(z^0-x^0)\vec\nabla_z{1\over|\vec z-\vec x|}
,
\end{eqnarray}
from which we find the dressed field
\begin{equation}\psi_c(x)=
\exp\left\{
ie{\partial_iA_i\over \nabla^2}(x)
\right\}\psi(x) .
\label{eq:Coul}
\end{equation}

\begin{figure}[t]
\vspace{9pt}
\hbox{\epsfxsize=6.49cm
\epsfbox{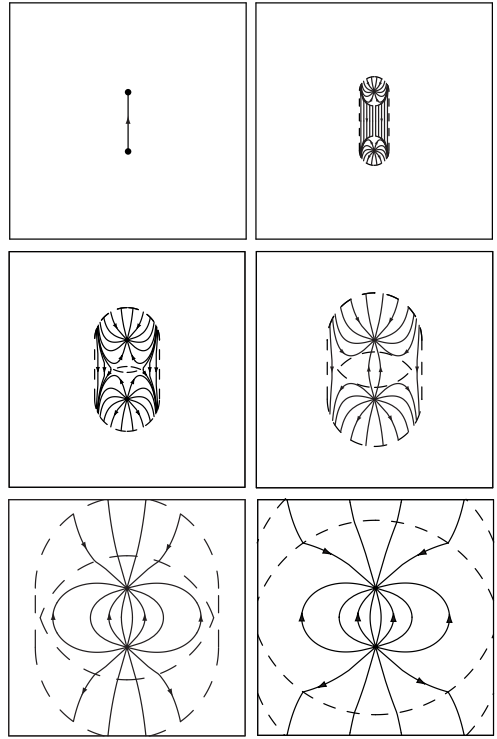}
}
\caption{Six frames\protect\cite{www} of the time evolution of a electric field initially 
located in the straight
line joining two opposite (static) charges. The dashed lines show the extension of the
region where the radiation field from the decay of the string is present.
Only the dipole field survives if we wait long enough.}
\label{fig:nicolas}
\end{figure}

Example~({\em i}) is not well suited to describe physical charges. First of all
it corresponds to a very singular field configuration. Second, 
its path dependence is difficult
to interpret and, finally, it has been shown that this field configuration is unstable~\cite{Shab}.
Essentially what happens is that the charge generates a Coulombic field and the string radiates away
to infinity. Fig.~\ref{fig:nicolas} shows the time evolution of a similar situation: the 
electric field is initially concentrated in the straight line
joining two charges in such a way that Gau\ss's law holds. Again the string radiates away and
only the dipole field generated by the charges remains in the far future\cite{www}.  

In contrast, example~({\em ii}) has very nice properties: ({\em a}) it is stable;
({\em b}) it is possible to factor out the path dependence in $\psi_\Gamma$ giving 
$\psi_c$, i.e.,
$\psi_\Gamma=N_\Gamma \psi_c $ where $N_\Gamma$ is a gauge invariant, but
path dependent, factor; ({\em c}) using the fundamental commutation relations
one has
\begin{equation}
[\vec E (\vec x),\psi_c(\vec y)]= 
{e\over4\pi}{\vec x-\vec y\over|\vec x-\vec y|^3}
\psi_c(\vec y)
\,,
\end{equation}
where we recognize the factor before $\psi_c$ as the electric Coulomb field.
This immediately suggests that example~({\em ii}) is the right dressing for a static charge.

One can generalize this dressing to the case of a charge moving with arbitrary velocity
$\vec v$
\begin{equation}
\psi_v=\exp\left\{
ie{g^{\mu\nu}\!-(\eta+ v )^\mu(\eta- v)^\nu\over\partial^2\!-
(\eta\cdot\partial)^2+(
v\cdot\partial)^2}\partial_\nu A_\mu
\right\}
\psi ,
\label{eq:boos}
\end{equation}
where $v=(0,\vec v)$, and $\eta=(1,\vec 0)$; ({\em d}) One can perform perturbative calculations
using these dressings. We shall do this in the next section where we also show that the 
(one loop)
mass shell renormalized dressed electron
propagator, 
\begin{equation}
i S_v(p)=\int d^4x\, \exp\{i p\cdot x\}\;
\langle0|\psi_v(x)\bar\psi_v(0)|0\rangle, 
\end{equation}
is infra-red finite 
provided $p=m\gamma(1,\vec v)$\cite{slow,fast}, 
as was already predicted in Ref.\cite{LaMcMu96a}. As far as we know,
there are only two other gauges with an infra-red finite charge propagator in the mass shell scheme:
the Yennie gauge and the Coulomb gauge. The latter is a particular case of our approach 
($\vec v\to \vec 0$). The infra-red finiteness of $S_v(p)$ can be understood as a consequence
of having the charge dressed with the asymptotic electromagnetic
field of a classical charge moving with velocity $\vec v$. This is a 
boosted Coulomb field. Since the
infra-red behaviour is related to its slow fall-off, one would expect that the same dressing 
should lead to infra-red finite results also for scalar QED.
This has been verified explicitly. The lack of infra-red divergences in the propagator
of the dressed charge is a necessary and highly non-trivial
requirement for the construction of an asymptotic
state with sharp momentum that can be interpreted as a (single) physical charge.

Before closing this section, we would like to comment on some subtleties 
associated with charged states.
Note that $\psi_v$ is both non local and non covariant, which one might regard as a problem.
However, it can be proved that these are {\em unavoidable} features of {\em any} 
operator that creates
charged {\em physical} states out of the vacuum\cite{LaMcMu96a,non-loc,non-cov}. 
Note also that Eq.(\ref{eq:boos})
is not just a Lorentz boost of Eq.(\ref{eq:Coul}). This is a consequence of the lack of covariance
and locality necessarily associated with charged states\cite{LaMcMu96a}.

\subsection{QCD}

All the properties that have been discussed in the previous section go through to QCD 
in perturbation theory. It is also possible to define dressed gluon fields
perturbatively. A new reason for introducing dressings in non-abelian gauge theories is that
the colour charge is only well-defined on gauge invariant states such as a dressed 
quark\cite{LaMcMu96a}. 
However, it has been shown that beyond perturbation theory the Gribov ambiguity
obstructs the construction of dressings\cite{LaMcMu96a}. As a result, one cannot obtain any
true observable out of a single lagrangian (quark) field.  Therefore, our approach explains 
confinement in the sense that one cannot construct an asymptotic quark field.

\section{THE DRESSED ELECTRON/QUARK PROPAGATOR}

\begin{figure}[htb]
\vspace{9pt}
\leftline{
\hbox{\epsfxsize=6.50cm
\epsfbox{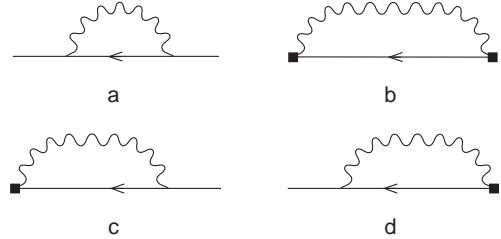}
}
}
\caption{The diagrams which yield the one loop dressed propagator, $S_v(p)$. The vertices
coming from the perturbative expansion of the dressing are denoted by a black box}
\label{fig:diagrams}
\end{figure}

The Feynman diagrams for the one loop dressed propagator, $S_v(p)$, are shown in
Fig.\ref{fig:diagrams}. 
The diagram
of Fig.\ref{fig:diagrams}.a gives the standard self~energy. The other three diagrams,
Figs.\ref{fig:diagrams}.b---d, contain a new vertex (the black box). The corresponding
Feynman rule, which can be easily obtained 
from the perturbative expansion of Eq.(\ref{eq:boos}), reads
\begin{equation}
e{(\eta+v)_\mu(\eta-v)_\rho-g_{\mu\rho}\over
k^2-(k\cdot\eta)^2+(k\cdot v)^2}k^\rho,
\end{equation}
where $k$ ($\mu$) is the momentum (Lorentz index) of the incoming photon.
We can now proceed in two ways: ({\em i}) compute the diagrams of Figs.\ref{fig:diagrams}.a---d
in a Feynman gauge and check that the result (before integrating the loop momentum)
is independent of the gauge parameter; ({\em ii}) use the so called {\em dressing gauge},
in which the dressing phase is 1, i.e., $\psi_v=\psi$ and compute only the diagram of 
Fig.\ref{fig:diagrams}.a. In the dressing gauge the photon propagator is
\begin{eqnarray}
&&
{1\over k^2}
\left\{-g_{\mu\nu}+
{k^2-[k\cdot(\eta-v)]^2\gamma^{-2}\over
[k^2-(k\cdot\eta)^2+(k\cdot v)^2]^2}\;k_\mu k_\nu\right.\nonumber\\
&&\left.
-{k\cdot(\eta-v)\over k^2-(k\cdot\eta)^2+(k\cdot v)^2}\;
k_{(\mu}\; (\eta+v)_{\nu)}\right\}.
\label{eq:prop}
\end{eqnarray}
Note that in the limit $\vec v\to\vec0$ this reduces to the propagator in Coulomb gauge.
We have explicitly checked that the two procedures give the same loop momentum integral.

To integrate the loop momentum we have chosen to work in dimensional regularization
with a space-time dimension $D=4-2\epsilon$.
This regularizes both the ultra-violet and infra-red divergences. In particular, the
latter show up as $\int_0^1 du\, u^{D-5}\sim 1/\epsilon$, where $u$ is a Feynman parameter.

\eject

\subsection{Ultra-violet divergences}

The ultra-violet divergent part of the electron self-energy has the following structure
\begin{eqnarray}
\Sigma_{UV}\sim &{\displaystyle{1\over\epsilon}}&\Bigl\{
-3m+(p\hspace{-0.45em}/-m){\cal F}_1(\vec v) \nonumber \\ &+&
2  [p\cdot v\eta\hspace{-0.45em}/
-p\cdot\eta v\hspace{-0.45em}/] {\cal F}_2(\vec v)
\Bigr\} ,
\label{eq:UVpiece}
\end{eqnarray}
where ${\cal F}_1$ and ${\cal F}_2$ do not depend on the external momentum, $p$. The last term
in Eq.(\ref{eq:UVpiece}) seems to endanger the multiplicative renormalization of the propagator.
However, one can check that Eq.(\ref{eq:UVpiece}) can be cancelled by introducing the standard
mass shift, $m\to m-\delta m$, and the 
following multiplicative 
{\em matrix} renormalization for the electron
\begin{equation}
\psi^{({\rm bare})}_v=\sqrt{Z_2} \exp\left\{
-i { Z'\over Z_2}\sigma^{\mu\nu}\eta_\mu v_\nu \right\}\psi_v,
\end{equation}
which is reminiscent of a naive Lorentz boost upon a fermion.

\subsection{Renormalization}

To actually compute $\delta m$, $Z'$ and $Z_2=1+\delta Z_2$ it is convenient to write
the renormalized self energy as
\begin{equation}
-i\Sigma=m\alpha+p\hspace{-0.45em}/\beta+p\cdot\eta \eta
\hspace{-0.45em}/\delta +m v\hspace{-0.45em}/ \varepsilon,
\end{equation}
where
$\alpha$, $\beta$, $\delta$  and $\varepsilon$ depend upon
$p^2$, $p\cdot\eta$,
$p\cdot v$  and $v$ and they also contain the counterterms  $\delta m$, $Z'$ and $\delta Z_2$.

We recall that in the mass shell scheme one has to impose the following 
two conditions: ({\em i}) The propagator has a simple pole at $m$, i.e.,
$m$ is the {\em physical} mass of the fermion;
({\em ii}) the residue of the propagator at $m$ is unity. From these two conditions
it must be possible to determine $\delta m$, $Z'$ and $\delta Z_2$.
Condition ({\em i}) implies that the renormalized $\Sigma$ must obey
\begin{equation}
\tilde\alpha +\tilde\beta+{(p\cdot\eta)^2\over m^2}\tilde\delta+
{p\cdot v\over m}\tilde\varepsilon=0,
\label{eq:cond1}
\end{equation}
where the tildes signify that we put the momentum $p^2$ on shell: $p^2=m^2$. Note
that $Z'$ and $\delta Z_2$ do not enter in~(\ref{eq:cond1}) so just $\delta m$ is determined.
We find
\begin{equation}
\delta m={e^2\over(4\pi)^2}\left(
{3\over\epsilon}+4\right).
\end{equation}
It is important to emphasize that this is the standard result for the mass shift and that
it solves Eq.(\ref{eq:cond1}) for arbitrary $p\cdot\eta$, $p\cdot v$ and $v$.

Condition ({\em ii}) can only be satisfied if $p=m\gamma(\eta+v)=m\gamma(1,\vec v)$,
which is precisely the momentum of the real electron moving with velocity $\vec v$. In
addition we need that
\begin{eqnarray}
0&=& 2 m^2 \bar\Delta +\bar\beta -\bar\delta\nonumber\\
0&=&\gamma\left(2m^2\bar\Delta+\bar\beta\right)-\bar\varepsilon\nonumber\\
0&=& 2 m^2 \bar\Delta +\bar\alpha +2\bar\beta,
\label{eq:cond2}
\end{eqnarray}
where
\begin{equation}
\Delta={\partial\alpha\over\partial p^2}+
{\partial\beta\over\partial p^2}+{(p\cdot\eta)^2\over m^2}
{\partial\delta\over\partial p^2}+{p\cdot v\over m}
{\partial\varepsilon\over\partial p^2}
\end{equation}
and the bar upon functions denotes that they must be computed at $p=m\gamma(1,\vec v)$.
If we now explicitly separate out the contributions of $Z'$ and $\delta Z_2$ from the
rest (to which we give a subscript R) then we find that (\ref{eq:cond2}) can be rewritten
as
\begin{equation}
\begin{array}{lcrcl}
-i\delta Z_2\negsp&+&\negsp2\vec v^2 i Z'\negsp&=&\negsp 2 m^2 \bar\Delta +
\bar\beta_R -\bar\delta_R\\
-i\delta Z_2\negsp&+&\negsp2 i  Z'\negsp&=&\negsp\gamma\left(2m^2\bar\Delta+
\bar\beta_R\right)-\bar\varepsilon_R\\
-i \delta Z_2\negsp&&\negsp\negsp&=&\negsp 2 m^2 \bar\Delta +
\bar\alpha_R +2\bar\beta_R.
\end{array}
\end{equation}
Since we have three equations and two unknowns ($Z'$ and $\delta Z_2$) one might worry
that perhaps no solution exists. However, a unique solution exists for our choice of mass shell.
It reads
\begin{eqnarray}
Z'&=&{1\over2i}[\gamma^2\bar\delta_R-\gamma\bar\epsilon_R]\nonumber\\
\delta Z_2&=&-{1\over i}[\bar\alpha_R+2\bar\beta_R+2m^2 \bar\Delta].
\end{eqnarray}
As is the case for the standard fermion propagator, the infra-red singularities can only enter
through derivatives with respect to $p^2$. Hence, $Z'$ is infra-red finite and
infra-red divergences can only arise in
$\delta Z_2$ through $\bar\Delta$. The total infra-red divergent contribution to $\bar\Delta$ is
\begin{eqnarray}
\bar\Delta_{{\rm IR}}\negsp&\sim&\negsp\!\!\int_0^1 du\, u^{D-5}
\left\{
-2+2\int_0^1 \!{dx \over\sqrt{1-x}\sqrt{1-\vec v^2 x}}\right.\nonumber\\
\negsp&\times&\negsp(1+\vec v^2-2\vec v^2 x)
\nonumber\\
\negsp&-&\negsp\left.
\gamma^{-2}\int_0^1  \! {dx\, x \over\sqrt{1-x}\sqrt{1-\vec v^2 x}}
{3+\vec v^2-2\vec v^2 x\over2(1-\vec v^2 x )}
\right\}\nonumber\\
\negsp&=&\negsp0,
\end{eqnarray}
and no infra-red divergence arises in the mass shell renormalize
propagator of the dressed electron. The full expressions for $Z'$ and $Z_2$ can
be found in Ref.\cite{fast}.
As we have already mentioned, one can also repeat the calculation in scalar QED
where the algebra is not so heavy. Again one can renormalize the propagator of
the dressed scalar electron, defined as in~(\ref{eq:boos}) replacing $\psi$ by the
scalar field. In this case, no $Z'$ is needed and we can get rid of the ultra-violet
divergences through the usual counterterms $Z_2$ and $\delta m$. Again,
no infra-red divergence arise. The consistency of these calculations with our
expectations is compelling evidence for the validity of
this approach.

\section{SUMMARY}

To conclude we note that
\begin{itemize}
\item Any description of a physical charge must be gauge invariant. Gau\ss's law implies
an intimate link between charges and a chromo-(electro-)~magnetic cloud.
\item Not all gauge invariant descriptions are physically relevant. One needs to find
the right ones.
\item In QCD there is no such description of a single quark outside of perturbation
theory. This sets the limits of the constituent quark model and fixes when jets start to
hadronise.
\item The perturbative tests reported here all worked. They are now being extended to vertex
studies.
\item Phenomenologically, the main question is: at what scale does the Gribov ambiguity
prevent any description of a quark from being stable?
\end{itemize}

\section*{Acknowledgments}
EB \& BF were supported by 
CICYT research project AEN95-0815 and  
ML by AEN95-0882. NR was supported 
by a grant from the region Rh\^one-Alpes.

\end{document}